# A HUMAN-CENTERED APPROACH TO REFRAMING JOB SATISFACTION IN THE BIM-ENABLED CONSTRUCTION INDUSTRY

**Sharareh Mirzaei**
***Gerald May Dept. of Civil, Construction and Environmental Engineering, University of New Mexico;***
*shmirzaei@unm.edu*

**Stephanie Bunt, Assistant Professor,**
***School of Architecture and Planning, University of New Mexico****.*
*sbunt@unm.edu*

**Susan M Bogus,**
***Professor, Gerald May Dept. of Civil, Construction and Environmental Engineering, University of New Mexico.***
*sbogus@unm.edu*

## ABSTRACT:

As the construction industry undergoes rapid digital transformation, ensuring that new technologies enhance rather than hinder human experience has become essential. The inclusion of Building Information Modeling (BIM) plays a central role in this shift, yet its influence on job satisfaction remains underexplored. In response, this study developed a human-centered measurement model for evaluating job satisfaction in BIM work environments by adapting Hackman and Oldham's Job Characteristics Model for the architecture, engineering, and construction (AEC) industry to create a survey that captured industry perspectives on BIM use and job satisfaction. The model uses Partial Least Squares Structural Equation Modeling to analyze the survey results and identify what dimensions of BIM-related work affect job satisfaction. While it was hypothesized that BIM use increases job satisfaction, the results show that only some dimensions of BIM use positively impact BIM job satisfaction; the use of BIM does not guarantee an increase in overall job satisfaction. Additionally, more frequent BIM use was not associated with higher satisfaction levels. These findings suggest that in the AEC industry, sustainable job satisfaction depends less on technological autonomy and more on human-centric factors—particularly collaboration and meaningful engagement within digital workflows.

## 1. INTRODUCTION

Building Information Modeling (BIM) is a transformative force in the architecture, engineering, and construction (AEC) industry, driving significant gains in productivity, data accuracy, and project delivery efficiency. By enabling real-time information exchange, reducing communication bottlenecks, automating routine processes, and restructuring coordination practices, BIM reshapes how projects are conceived and executed, from early design phases to facility operations and maintenance (Hallén et al., 2023).

While prior research has centered on BIM's technical potential, this study shifts its focus toward the human impact of BIM by developing a framework for understanding job satisfaction in digital AEC workplaces. The digital adoption process should not be viewed as a one-time event, but rather as an ongoing transformation in how work is organized and performed. Continuous post-adoption evaluation is essential to ensure that both the technical and human dimensions of change are addressed (Greenhalgh et al., 2017).
Understanding how digital systems reshape workers' engagement is crucial not only for sustaining productivity gains, but also for maintaining job satisfaction in technologically mediated environments (Bolli & Pusterla, 2022). Given the AEC industry's reliance on human labor (Dainty et al., 2007) and its long-standing issues with job dissatisfaction (CDC, 2016; Holistic Healthcare Group, 2020; Tijani et al., 2021), studying job satisfaction has become increasingly important, especially as emerging digital technologies introduce new complexities and demands into the workplace.

This study addresses a gap in current research by examining job satisfaction (JS) within BIM-enabled work environments through the development of a unique BIM Job Satisfaction (BIM-JS) Measurement Model. The model defines BIM-JS by integrating Hackman and Oldham's Job Characteristics Model (JCM), key job satisfaction indicators identified through an extensive literature review, and persistent pain points specific to the AEC industry. This approach offers a human-centered perspective on the digital workplace in AEC. There are three JCM indicators used in the model. The first is experienced meaningfulness, defined as the extent to which BIM-enabled work supports skill variety, task identity, and task significance (Hackman, 1971). The second involves feedback loops, referring to the extent to which BIM enables clear and timely performance reviews and evaluation (Hackman, 1971). The third is autonomy, which refers to the degree to which professionals can make independent decisions when using BIM tools. While the JCM provides a strong foundation for analyzing work satisfaction, its original formulation does not fully reflect the realities of digital, collaborative environments. To address this gap, the proposed BIM-JS Measurement Model extends the three JCM indicators by introducing two additional factors: collaboration, a strong predictor of job satisfaction in digital contexts (Wong et al., 2014), and pain-point improvement, which captures the social and experiential aspects of BIM-enabled work. Collaboration represents the quality of interdisciplinary engagement fostered by BIM, while pain-point improvement reflects technology's role in mitigating long-standing coordination inefficiencies common in AEC projects. Together with meaningfulness, autonomy, and feedback, these factors form a human-centered framework for evaluating job satisfaction in digital construction settings.

The BIM-JS Measurement Model is used to examine the first research question (RQ1), which asks which of the JS factors (meaningfulness, collaboration, autonomy, feedback, and pain-point improvement) significantly and positively contribute to the formation of the BIM Job Satisfaction construct? The second research question (RQ2) asks if BIM-related job satisfaction can be used to predict overall job satisfaction? After creating the BIM-JS Measurement Model, the structural relationship between BIM-JS and Overall-JS can be assessed. Although BIM may improve JS, we recognize that not all professionals use BIM the same amount, leading to the third research question (RQ3), which asks whether the frequency of BIM use influences BIM-JS or Overall-JS? To address these questions, hypotheses about the JS factors were evaluated using Partial Least Squares Structural Equation Modeling, and a Confirmatory Composite Analysis was used to validate the structure, reliability, and appropriateness of composite constructs in the BIM-JS Measurement Model, ensuring that the indicators and their relationships to constructs are theoretically and statistically justifiable. By centering on human experience, this study offers a novel perspective on BIM-JS in the AEC industry.

## 2. LITERATURE REVIEW

### 2.1. BIM Definitions and Benefits

BIM has emerged as a transformative digital innovation in the AEC industry, offering a collaborative and integrated approach to managing the lifecycle of built assets. Although definitions vary by context, BIM is broadly recognized as a 3D model-based methodology for planning, designing, constructing, and operating infrastructure (Muta et al., 2025). It also functions as a comprehensive system for managing design and project data (Volk et al., 2014), and as a collaborative framework that integrates technological, managerial, and human factors to enhance project execution and facility management (Oraee et al., 2017).

At its core, BIM consolidates multidisciplinary data within a unified digital environment, improving accuracy, reducing errors, and increasing transparency throughout the project lifecycle (Hallén et al., 2023). Its growing adoption reflects broader digitalization trends across the construction sector, with documented improvements in productivity, decision-making, and operational efficiency (Azhar, 2011). Through synchronized, real-time information exchange, BIM enhances process efficiency, fosters stakeholder accountability, and strengthens team motivation (Lee et al., 2023). Cloud-based BIM platforms further support coordination, with studies showing significant improvements in project outcomes (Bakker et al., 2005). When paired with structured change management and ongoing training, organizations report productivity gains ranging from 70% to 240% (Oakland & Tanner, 2007; Poirier et al., 2015).

The human-centered benefits of BIM are particularly evident in its capacity to streamline collaboration among stakeholders (Villena Manzanares et al., 2024). Its digital environment supports iterative updates and centralized feedback, facilitating effective coordination across disciplines, including engineers, supervisors, and contractors (Azhar, 2011; Bandura, 1977; Guo et al., 2019). One of the most impactful features of BIM is its advanced visualization capability; interactive 3D models help both technical and non-technical users understand design intent, enabling earlier feedback, higher design quality, and reduced rework (Chu et al., 2018). Beyond coordination, BIM also supports performance simulation, allowing teams to model outcomes such as energy efficiency and structural behavior, which helps anticipate risks and improves confidence in cost, scheduling, and feasibility assessments (Villena Manzanares et al., 2024).

Technologically, BIM provides a foundation for automation, real-time data exchange, and AI-powered analytics. Its cloud-based architecture ensures continuous access to up-to-date project data, supporting more agile and responsive management. Furthermore, BIM's compatibility with emerging technologies, such as digital twins and the Internet of Things (IoT), enhances its value by enabling predictive maintenance, optimizing building performance, and supporting intelligent facility management (Love et al., 2014).

However, despite the considerable promise of BIM's technical capabilities, their effective realization in practice remains uneven. Implementation efforts are often impeded not solely by technical limitations but by a broader set of challenges. Organizational inertia, workflow disruptions, and resistance to change among stakeholders continue to present substantial barriers to successful integration. The following section explores these constraints in depth, examining the technological, human, and institutional factors that influence the adoption and effective use of BIM across the AEC industry.

## 2.2. BIM User Resistance

Human-related challenges remain some of the most persistent and underexamined barriers to effective BIM implementation. In a human-intensive industry, successful technological transitions depend as much on user readiness and acceptance as they do on the capabilities of the tools themselves.

Resistance to change is the most widely documented human obstacle. Contributing factors include behavioral inertia, limited organizational support, and concerns over productivity disruptions (Evans & Britt, 2023). Psychological and cognitive barriers such as status quo bias, uncertainty about new technologies, and emotional attachment to legacy systems like CAD, further complicate adoption (Klein et al., 2022). Many professionals weigh the cognitive effort of learning new tools against the perceived loss of competence or investment in familiar systems. Addressing these issues requires more than technical solutions. Effective change management involves leadership commitment, consistent communication, and the inclusion of end-users in decision-making processes (MacLoughlin & Hayes, 2019). When users are engaged and supported through training, resistance often diminishes, and long-term benefits become clearer. Training is important and studies recently encourage interactive and practice-based formats—not only traditional or rigid formats, to better reflect real workflows and improve engagement (Akbari et al., 2025).

A related concern is the persistent skills gap across the AEC workforce. New graduates frequently lack adequate BIM training, while experienced professionals often require reskilling to keep pace with evolving digital workflows (Suprun et al., 2019). Without targeted education and ongoing professional development, organizations risk underutilizing BIM and slowing broader adoption (Arayici et al., 2011; Qin et al., 2020). Closing this gap requires both curriculum reform and institutional support for continuous learning. Beyond individual challenges, organizational culture significantly shapes BIM outcomes. Cultural resistance often stems from a lack of preparedness for digital transformation. Leadership attitudes, openness to innovation, and tolerance for change all influence adoption success (Alankarage et al., 2023). When employees are excluded from planning or perceive BIM as imposed without purpose, skepticism and disengagement can undermine implementation (Arayici et al., 2011).

Despite frequent emphasis on technical and organizational factors, the human dimension is equally critical. BIM alters workflows, redefines responsibilities, and demands new competencies—changes that directly affect job satisfaction and professional identity. Adoption, therefore, involves not only technological adjustment but also cultural and psychological transformation (Arayici et al., 2011; Poirier et al., 2017; Suprun et al., 2019).

While recent studies have explored human aspects, most focus narrowly on the initial adoption phase. Post-adoption experiences remain underexplored, particularly in relation to job satisfaction—an essential yet overlooked lens for evaluating BIM's long-term impact on the construction workforce.

## 2.3. Job Satisfaction & Measurements

Job satisfaction is one of the most widely studied topics in organizational research. It reflects how employees feel and think about their work. Traditionally, job satisfaction has been defined as a positive emotional state that results from evaluating one's job or work experiences (Sonnenfeld, 1985). It includes how people respond to their tasks, work environment, and how well their job aligns with personal goals. Earlier research focused on external factors like pay, job security, and working conditions. However, more recent studies emphasize internal factors such as autonomy, purpose, opportunities for growth, and meaningful work (Ćulibrk et al., 2018; Sang et al., 2009). These intrinsic elements are especially important in modern, knowledge-based jobs where people value personal development and impact. Job satisfaction has been linked to many positive outcomes, including higher productivity, lower turnover, stronger commitment, and greater innovation (Somvir & Kaushik, 2012; Villena Manzanares et al., 2024).

## 2.4. Job Characteristics Model

To systematically understand how job characteristics influence satisfaction and motivation, Hackman and Oldham developed the JCM in the 1970s, a framework that remains widely applied in contemporary job design research. The model identifies five core job factors: skill variety, task identity, task significance, autonomy, and feedback. Each of which contributes to three critical psychological states: experienced meaningfulness of work, responsibility for outcomes, and knowledge of results of activity (J. Hackman, 1974; J. R. Hackman, 1980). When these psychological states are activated, employees are more likely to experience intrinsic motivation and higher job satisfaction.

The feeling that one's job is meaningful arises when individuals perceive their tasks as significant and valuable (Hackman, 1971). This sense of meaningfulness is commonly shaped by three core job characteristics: skill variety, task identity, and task significance. Skill variety refers to the presence of diverse work activities and the opportunity to apply a range of skills to accomplish tasks. Task identity is defined as the ability to complete an entire and clearly identifiable piece of work. Task significance reflects the perceived impact of one's job on others and its contribution to organizational success (Hackman, 1971). When these elements are present, employees are more likely to find their work engaging and personally fulfilling. In the context of BIM, job satisfaction may be enhanced by increasing employees' sense of meaningfulness in their work. This leads to the first hypothesis:

- H1a: Meaningfulness positively and significantly predicts BIM-JS.

Beyond a sense of meaningfulness, JCM measures feedback and autonomy through the process. Feedback relates to the clarity and immediacy of information employees receive about their job performance. When workers can observe the results of their efforts through direct task-related cues, it helps them evaluate their effectiveness, make necessary improvements, and enhance confidence (Belletier et al., 2021). Autonomy, on the other hand, refers to the degree of independence an employee has in making decisions on their work and feels responsible for the outcome. Jobs that offer high autonomy tend to foster a stronger sense of accountability, as individuals feel their input directly affects outcomes (Hackman, 1971). These factors play a crucial role in creating motivating work environments that support job satisfaction and performance (Hackman, 1971). Therefore, this study examines these factors of job satisfaction through the following hypotheses:

- H1b: Feedback positively and significantly predicts BIM-JS.
- H1c: Autonomy positively and significantly predicts BIM-JS.

Numerous studies have validated the JCM's utility in various sectors, including manufacturing, healthcare, education, and construction. For instance, in environments where task repetition and hierarchical control dominate, JCM helps to redesign tasks to improve worker engagement by increasing autonomy and clarifying the importance of tasks (Sun et al., 2022). In technology-supported roles, the JCM has been applied to evaluate levels of job satisfaction. Real-time feedback tools, such as wearable devices and performance dashboards, align well with the model's feedback dimension, enhancing workers' responsiveness and confidence (Belletier et al., 2021). Additionally, JCM has been used to evaluate job design in knowledge-intensive industries, where the complexity of tasks and autonomy in problem-solving are key contributors to satisfaction. These findings demonstrate that JCM remains a relevant and flexible model for studying workplace motivation and satisfaction, particularly in environments that seek to balance structure and innovation.

However, as workplaces become more digitized and reliant on information systems, scholars have begun to question whether the JCM adequately captures the dynamics of technology-mediated work. In particular, digital tools like BIM, Enterprise Resource Planning (ERP) systems and Artificial Intelligence (AI) platforms introduce new forms of task interaction, decision-making, and feedback that differ substantially from the work environments originally described in the JCM. For example, Morris and Venkatesh (2010) found that the JCM's five core factors explained only 47% of the variance in job satisfaction among ERP system users, suggesting the presence of other unaccounted-for variables such as system usability, interface quality, and user trust. These findings highlight the limitations of relying solely on traditional job design frameworks when assessing job satisfaction in digitally transformed workplaces. Accordingly, this study expands the JCM to capture job satisfaction factors specific to BIM-enabled work environments within the AEC industry. Recent critiques highlight that conventional job satisfaction models often overlook the role of digital systems in shaping modern work experiences. Key factors such as system usability, perceived usefulness, data accuracy, and platform reliability are now essential to satisfaction in digital roles, yet they are often missing from classical frameworks like the JCM (Murphy et al., 2012). Despite being somewhat dated, the JCM remains among the most widely cited models in this field, focusing on how specific job features impact motivation and satisfaction through various psychological mechanisms.

## 2.5. Beyond JCM

While the JCM provides a foundational framework for analyzing job satisfaction across a wide range of roles and industries (Hackman & Oldham, 1976), many scholars have argued that its generality limits its ability to fully capture the complexity of contemporary work environments. Consequently, numerous studies have extended or adapted the JCM to better align with specific domains or technological contexts. More recent research in high-tech and project-based settings highlights additional domain-sensitive factors, including system usability, technological dynamism, and interdisciplinary collaboration (Morgeson et al., 2006; Bayo-Moriones et al., 2010). These developments underscore that, although the JCM identifies essential psychological dimensions of work, it must be combined with domain-specific considerations to accurately reflect the factors shaping job satisfaction in modern, technology-intensive environments. To contribute to this ongoing effort to expand the model, the present study explores how job satisfaction is measured in related domain-specific research such as, Information System, to identify which factors are most emphasized as critical to the employee experience.

### 2.5.1. Collaboration

A review of domain-specific literature reveals that collaboration frequently emerges as one of the most critical factors used to measure job satisfaction, particularly in digitally intensive and project-based environments such as AEC. Across numerous studies, collaboration is a critical driver of job satisfaction, as it fosters shared responsibility, mutual support, and a sense of collective achievement (Pedrycz et al., 2011). Workplace dynamics shaped by effective teamwork can significantly influence how individuals perceive their roles and overall satisfaction (Locke, 1969). This is a complex issue, particularly in the modern workplace, where individual job satisfaction is closely linked to both team performance and overall team satisfaction (Pedrycz et al., 2011). In 1984, Goldstein and Rockart extended the JCM by introducing variables that account for relationships among coworkers. Their research highlighted that collaborative work significantly moderates the link between job characteristics and job satisfaction. For example, the satisfaction derived from a task can vary depending on whether it is performed individually or within a collaborative context (Wong et al., 2014). Collaboration and information exchange remain persistent challenges in the construction industry. These challenges largely stem from the fragmented and decentralized nature of data management within the sector (Saka & Chan, 2021). In this context, Cloud-based BIM presents a transformative and cooperative environment by offering a centralized digital platform that promotes seamless communication and coordinated project workflows (Wong et al., 2014; Souza et al., 2023). Given the emphasis on collaboration as a key factor in job satisfaction and the importance of that in AEC, we extend the existing model with the following hypothesis:

- H1d: Collaboration positively and significantly predicts BIM-JS.

To define the final key factor contributing to job satisfaction in the AEC industry, it is essential to study how BIM affects long-lasting dissatisfaction. Despite sustained efforts to boost productivity, improve project outcomes, and enhance team collaboration, widespread dissatisfaction persists. These issues endure even in the face of advancing technologies, indicating that digital tools are insufficient unless they address the deeper causes of dissatisfaction. Accordingly, the following section critically explores the dissatisfaction that has long shaped the AEC work environment. Understanding these persistent frustrations is essential to evaluating whether—and to what extent—BIM can alleviate them and thereby enhance job satisfaction among construction professionals.

### 2.5.2. AEC Persistent Pain Points

Despite growing awareness, 73% of construction employees believe their employers offer inadequate mental health support, and 70% report experiencing stress, anxiety, or depression (Mostert et al., 2011). These figures underscore an urgent need for professional interventions to improve well-being in the AEC sector. The industry's fragmented structure, project-specific workflows, labor-intensive operations, high-risk business environment, and resistance to automation contribute to stressful and often hazardous working conditions (Hu & Panthi, 2018). Compounding these issues is the entrenched culture of extended work hours, which erodes work-life balance and heightens occupational stress (Hu & Panthi, 2018). Such conditions are strongly linked to lower job satisfaction, diminished organizational commitment, higher turnover, and reduced productivity (Cheung et al., 2022). Additionally, AEC projects are often awarded short notice in highly competitive environments, requiring the rapid formation of project teams. This volatility forces firms to manage fluctuating workloads without stable work volume assurance (Dainty & Loosemore, 2013), resulting in irregular schedules, tight deadlines, and elevated stress levels (Zheng & Wu, 2018). The industry's cyclical downturns and persistent uncertainty further intensify these pressures, as staffing reductions shift greater responsibility onto fewer workers (Hu & Panthi, 2018).

Also, on the individual level, construction workers experience elevated job expectations, limited task autonomy, and inadequate social support (Woje et al., 2023). These psychological stressors contribute to a range of negative outcomes, including increased absenteeism, high turnover, lower work quality, and diminished productivity, all of which increase costs and delay project completion (Woje et al., 2023). The Health and Safety Executive (HSE) attributes 80–90% of industrial accidents to personal issues and unmanaged stress (Jansen, 1986), while the European Agency for Safety and Health at Work estimates that stress causes roughly 50% of job-related absenteeism (Simmons & Simmons, 1997).

To assess the extent to which BIM has helped alleviate persistent pain points and contributed to job satisfaction, the following final hypothesis was formulated to complete the response to Research Question 1:

- H1e – Pain points relief positively and significantly predicts BIM-JS.

## 2.6. From BIM-JS to Overall-JS: Direct and Mediated Effects

Having identified the key BIM work factors, meaningfulness, autonomy, feedback, collaboration, and pain points, that shape the BIM-JS construct, the next step is to examine its broader impact on Overall-JS. This analysis helps determine the extent to which BIM-JS contributes to overall satisfaction compared to other established predictors of Overall-JS. To address this question, the following hypothesis is proposed:

- H2: BIM-JS positively and significantly predicts Overall-JS.

In addition to examining the direct relationship between BIM-JS and Overall-JS, this study investigates whether the frequency of BIM use acts as a mediating mechanism between the two. Specifically, it explores whether the extent of professionals' engagement with BIM tools influences job satisfaction outcomes. To assess this potential mediation effect, the following hypotheses are proposed:

- H3a (Mediation Path 1): Frequency of BIM use positively and significantly influences the BIM-JS.
- H3b (Mediation Path 2): Frequency of BIM use positively and significantly influences Overall-JS.
- H3c (Indirect Effect): Frequency of BIM use positively mediates the relationship between BIM-JS and Overall-JS. That is, the association is stronger when BIM is used more frequently.

Combining all hypotheses, Figure 1 illustrates the proposed BIM-JS Measurement Model.

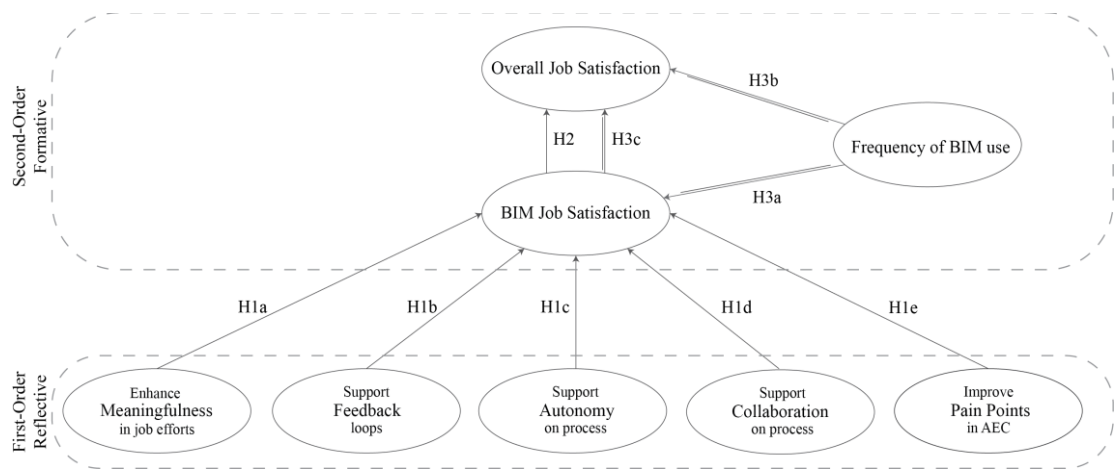

*Fig.1: The Conceptual BIM-JS Measurement Model*

## 3. METHODOLOGY

Following a comprehensive literature review, developing hypotheses, and the construction of the conceptual measurement model, a survey instrument was designed to collect AEC worker's perspectives on BIM in their workplace. To clearly illustrate the IRB approved research process, Fig. 2 presents the sequential workflow in this study and references the section numbers where each step is developed. The following sections detail the methodology employed in this paper.

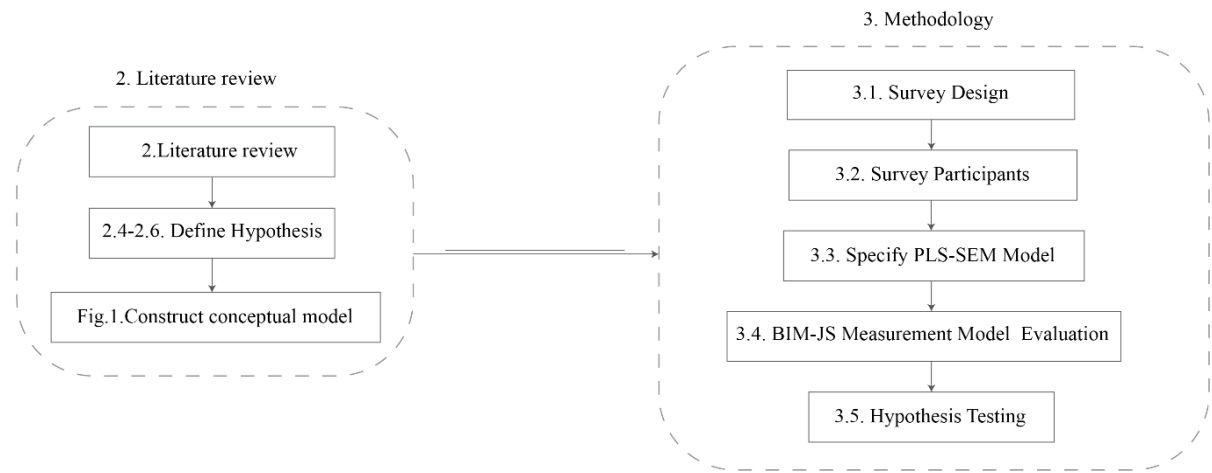

Fig. 2: Research Workflow

## 3.1. Survey Design

This study used a cross-sectional online survey to investigate job satisfaction and BIM use among professionals in the AEC industry. The survey instrument was designed to capture information on participants' demographic characteristics, professional background, level of BIM engagement, as well as overall job satisfaction and BIM-specific job satisfaction, to develop a model to measure BIM-JS.

The BIM-JS Measurement Model combined five JS factors to assess BIM-mediated work dynamics. First, the JCM was adapted to the AEC context, identifying core factors such as meaningfulness, autonomy, and feedback. Collaboration, as the fourth factor, was synthesized from a targeted literature review on technology-enabled work, drawing from both AEC and related domains. Lastly, the framework incorporated AEC-specific pain points reflecting persistent operational frictions that are known to negatively impact job satisfaction. In the survey, each factor included questions designed to capture participants' perceptions of that factor within a BIM-enabled workplace. Each was measured using a specific number of survey items, depending on its scope and complexity. For example, meaningfulness was assessed using seven items (MF1–MF7), reflecting the full range of sub-factors in Hackman and Oldham's JCM (i.e., skill variety, task identity, and task significance). In contrast, factors such as feedback (FB1) and autonomy (Aut1) were measured with single items due to their more narrowly defined operational scope.

Together, these components allow for the study of job satisfaction through various dimensions and provide a holistic perspective on BIM-JS. All survey items, except those in the demographic section, employed a 5-point Likert scale (1 = strongly disagree; 5 = strongly agree). The survey questions used in the data analysis are provided in Table 1.

Table 1: Survey items used in the BIM-JS Measurement Model.

| Code | | Question |
| --- | --- | --- |
| OJ1 | Overall-JS | In general, I enjoy doing my current job. |
| OJ2 | Overall-JS | I understand my responsibilities |
| OJ3 | Overall-JS | I enjoy the people I work with. |
| OJ4 | Overall-JS | My current job aligns with my career goals. |
| OJ5 | Overall-JS | I feel that what I am doing at my job is important. |
| MF1 | Meaningfulness | BIM enables me to engage in diverse project areas beyond my usual tasks. |

| Code | | Question |
|---|---|---|
| MF2 | Meaningfulness | BIM models and data in coordination meetings help me learn from contractors and apply diverse skills. |
| MF3 | Meaningfulness | BIM makes me feel more responsible for my tasks. |
| MF4 | Meaningfulness | BIM clarifies how my tasks fit into the broader project and the final result. |
| MF5 | Meaningfulness | BIM models and data help me feel that my work benefits the team. |
| MF6 | Meaningfulness | BIM models and data help me see my work's contribution to client value. |
| MF7 | Meaningfulness | BIM models and data help me see how my work supports larger goals, such as sustainability |
| Aut1 | Autonomy | BIM provides the information and tools needed to make independent task decisions. |
| FB1 | Feedback | BIM enables quick and constructive feedback from supervisors and team members. |
| Col1 | Collaboration | BIM models and data enhance internal team collaboration, such as sharing updates or solving issues together. |
| Col2 | Collaboration | BIM fosters collaboration across different teams, such as design, engineering, and construction. |
| Col3 | Collaboration | BIM improves communication and coordination between on-site and off-site teams. |
| Col4 | Collaboration | BIM makes my workflow easier and more organized. |
| PS1 | Pain point | BIM reduces manual drafting and simplifies design visualization. |
| PS2 | Pain point | BIM reduces the stress of tight deadlines by streamlining processes. |
| PS3 | Pain point | BIM models and data reduce mental stress by improving clarity and communication, and by reducing rework. |
| PS4 | Pain point | I have saved more time by using BIM. |

OJ = Overall Job Satisfaction, MF = Meaningfulness, Aut = Autonomy, FB = Feedback, Col = Collaboration, PS= Pain point Improvement

### 3.2. Survey Participants

After designing the survey and piloting the survey with five individuals, the final version was distributed to the study participants. Recruitment took place between May 2025 and August 2025 through LinkedIn, Autodesk forums, and Autodesk user groups, with a focus on individuals actively involved in BIM-enabled work. A total of 119 responses were collected. The data were screened for missing values, scale validity, and response quality. Following this screening and cleaning process, 86 responses were retained for analysis. The survey collected information about the gender, age, education level, BIM experience, professional roles, BIM usage in projects, and which BIM software they use most regularly. A summary of the descriptive characteristics of the respondents is shown in Table 2.

*Table 2: Descriptive data of participants*

| BIM Usage in Projects Category | Count | Percentage |
|---|---|---|
| Sample Size | 86 | 100% |
| **Gender Distribution** | | |
| Male | 58 | 67.4 |
| Female | 28 | 32.6 |
| **Age Groups** | | |
| Under 25 | 7 | 8.1 |
| 25-34 | 38 | 44.2 |
| 35-44 | 25 | 29.1 |
| 45-54 | 13 | 15.1 |
| 55+ | 3 | 3.5 |

| Education Levels | | |
|---|---|---|
| Under a bachelor's degree | 11 | 12.8 |
| Bachelor's degree | 33 | 38.4 |
| Above master's degree | 42 | 48.8 |
| **Work Experience** | | |
| Less than 1 year | 0 | 0 |
| 1-3 years | 15 | 17.4 |
| 4-6 years | 21 | 24.4 |
| 7-10 years | 16 | 18.6 |
| 10+ years | 34 | 39.5 |
| **BIM Experience** | | |
| Less than 1 year | 7 | 8.1 |
| 1-3 years | 15 | 17.4 |
| 4-6 years | 21 | 24.4 |
| 7-10 years | 19 | 22.1 |
| 10+ years | 24 | 27.9 |
| **Professional Roles** | | |
| Architect | 25 | 29.1 |
| Engineer | 20 | 23.3 |
| Project Manager | 9 | 10.5 |
| BIM Specialist | 23 | 26.7 |
| Contractor | 5 | 5.8 |
| Other Roles | 4 | 4.6 |
| **BIM Software** | | |
| Only Autodesk Revit | 40 | 44 |
| Autodesk Revit + Navisworks | 33 | 36 |
| Other software | 13 | 15 |
| **Frequency of BIM Use in Project** | | |
| Under 15% | 7 | 8 |
| 16-30% | 8 | 9 |
| 31-60% | 9 | 10 |
| Above 60% | 62 | 72 |

### 3.3. Specify PLS-SEM Model

In this step, the conceptual model developed in the literature review was converted into an empirical PLS-SEM model using the Two-Stage Hierarchical Component Modeling approach. First, a conceptual framework was built from prior literature to identify the main factors and formulate testable hypotheses. These hypotheses were then evaluated using PLS-SEM in SmartPLS, with additional diagnostics performed in the statistical analysis program, R. PLS-SEM was chosen because it is well-suited for small sample sizes, non-normal data, and formative constructs, and because the goal of this study is to predict and identify the key drivers of shaping BIM-related job satisfaction (Hair Jr et al., 2021; Henseler et al., 2016). While the PLS-SEM algorithm can involve multiple stages of regression estimation, the two-stage approach was used for this research as it is particularly appropriate for modeling multidimensional, higher-order constructs, like BIM-JS, where several distinct components combine to form the overall concept. Constructs such as satisfaction or happiness do not exist as standalone entities; they are inherently abstract and must be understood through their underlying dimensions or components, which together give the construct its full meaning. In addition, the growing use of this method in recent construction management studies reinforces its suitability for analyzing complex, digitally mediated, and human-centered constructs, further validating its application in this research context (Kineber et al., 2021; Mia et al., 2022; Villena Manzanares et al., 2024).

For this work, the two-stage PLS-SEM algorithm was used. In Stage 1, first-order factors, *meaningfulness*,

*collaboration*, and *pain points*, were specified as Mode A (reflective) composites; *autonomy* and *feedback* were single-indicator composites. In Stage 2, first-order latent scores served as formative indicators of the higher-order composite BIM-JS, which then predicted Overall-JS. BIM-JS was modeled as a formative second-order construct because its dimensions (meaningfulness, collaboration, autonomy, feedback, and pain-point improvement) represent distinct and non-interchangeable aspects of BIM-mediated work. Each dimension contributes uniquely to the overall construct, and removing any of them would alter the conceptual meaning of BIM-JS.

### 3.4. BIM-JS Measurement Model Evaluation

### 3.4.1. Model Evaluation

After specifying the model, it is necessary to assess the quality of measurements to ensure the model's reliability and validity (Henseler et al., 2016). For stage 1 (reflective measurement models), Confirmatory Composite Analysis (CCA) was employed to validate first-order reflective constructs. It involves assessing several key aspects: (1) indicator reliability, (2) internal consistency reliability and convergent validity, and (3) discriminant validity (AlNuaimi et al., 2021).

Indicator reliability was assessed by examining outer loadings, which were expected to exceed the recommended threshold of $\lambda > 0.708$ to confirm that each item of construct effectively captured its intended construct (Hair Jr et al., 2017). Internal consistency was evaluated using Composite Reliability (CR) to ensure that the items measured the same underlying concept, with CR values above 0.70 indicating acceptable reliability (Nunnally & Bernstein, 1994). Convergent validity was examined using Average Variance Extracted (AVE), where $AVE > 0.50$ demonstrates that the indicators of each factors share sufficient common variance (Fornell & Larcker, 1981). Discriminant validity was assessed using both the Fornell–Larcker criterion and the Heterotrait–Monotrait ratio (HTMT), with HTMT values below $HTMT < 0.85$ (strict) or $< 0.90$ (lenient) confirming factors are distinct (Henseler et al., 2015).

Following the CCA analysis in Stage 1, Stage 2 involved evaluating the formative second-order construct. First, multicollinearity was assessed to ensure that the formative indicators provided distinct information and were not excessively correlated, thereby allowing their individual effects on the results to be clearly distinguished. All outer Variance Inflation Factor (VIF) values were below the recommended threshold of 3.3, indicating no critical collinearity issues (Hair Jr, 2021). Next, the outer weights of the first-order dimensions were analyzed to assess their relative contribution to the higher-order BIM–JS composite. Because formative indicators require significance testing, bootstrapping with 5,000 subsamples was applied to evaluate the significance of the outer weights (and outer loadings when needed), using $p < 0.05$ as the criterion. Bootstrapping is essential for formative measurement models because it provides distribution-independent standard errors, ensuring that each dimension contributes meaningfully to the higher-order construct.

### 3.4.2. Structural Model Evaluation

After validating the measurement model, the structural model was assessed to evaluate whether the hypothesized relationships are supported and whether the model has explanatory and predictive power (AlNuaimi et al., 2021). Model fit was examined using the Standardized Root Mean Square Residual (SRMR); values below 0.1 indicate that the correlations implied by the model closely match the observed correlations, suggesting an acceptable global fit (Hu & Bentler, 1999). Predictive relevance was evaluated using Stone–Geisser's $Q^2$ obtained via blindfolding. $Q^2$ values above zero indicate that the model predicts the endogenous construct better than a baseline approach, with higher values reflecting stronger predictive relevance (Hair Jr et al., 2021). The model's explanatory power for job satisfaction was assessed using $R^2$, which indicates the proportion of variance explained by its predictors. Finally, Cohen's $f^2$ was reported to determine the contribution of each predictor to $R^2$; values of 0.02, 0.15, and 0.35 represent small, medium, and large effects, respectively (Cohen, 1988).

### 3.5. Hypothesis Testing

For the first research question (H1), path weights from each first-order construct to BIM-JS were examined. This tested whether dimensions such as collaboration, feedback, and autonomy significantly shaped satisfaction with BIM-related work. Loadings and outer weights were used to identify which factors had the strongest influence. The second hypothesis (H2) examined whether BIM-JS, as a higher-order construct, predicted overall job satisfaction. This involved evaluating the direct structural path from BIM-JS to overall-JS. Finally, the third hypothesis (H3) assessed whether the frequency of BIM use influenced overall job satisfaction, and whether this effect was mediated by BIM-JS. The model examined both direct and indirect paths, allowing for the identification of full or partial mediation effects. Together, these procedures enabled a thorough examination of the conceptual model and provided a foundation for interpreting the structural findings presented in the results.

## 4. RESULTS

This section reports the evaluation of the measurement and structural models of PLS-SEM, following the procedures described in the methodology. The results of hypothesis testing are then presented to assess support for the proposed hypotheses.

### 4.1. Model evaluation

The measurement model evaluation emphasized reliability, validity, and construct assessment. Consistent with the two-stage hierarchical component modeling approach, the reflective first-order constructs were assessed in Stage 1 using CCA and then in Stage 2 multicollinearity for the formative second-order construct (BIM-JS) was assessed.

#### 4.1.1. Stage 1: Reflective Measurement Model (CCA)

#### 4.1.1.1. Indicator Reliability

In this study, all first-order constructs were modeled as reflective (Mode A) composites within the PLS-SEM framework. The higher-order construct (BIM–JS) was assessed using the two-stage approach (Fincham et al., 2008). During the first stage, the reliability of each indicator was evaluated by examining its factor loading ($\lambda$).

To improve reliability without compromising content validity, indicators with loadings below the recommended threshold of 0.708 were removed. This was done gradually, removing only one low-loading item at a time to preserve as much of the construct's content as possible. The item with the weakest loading (MF7) was eliminated first. The model was then re-estimated, and additional items with low loadings (OJ2, OJ3, and Col3) were removed in successive rounds. After four rounds of refinement, all remaining items met the required loading criteria (Table 3), and the measurement model was considered acceptable for further assessments.

*Table 3: Stage-1 reflective loadings.*

| Construct | Loading $\lambda$ | Construct | Loading $\lambda$ |
|---|---|---|---|
| OJ1 | 0.879 | Col1 | 0.853 |
| OJ4 | 0.802 | Col2 | 0.837 |
| OJ5 | 0.884 | Col4 | 0.786 |
| MF1 | 0.756 | Aut1 | 1.000 |
| MF2 | 0.854 | FB1 | 1.000 |
| MF3 | 0.836 | PS1 | 0.808 |
| MF4 | 0.790 | PS2 | 0.721 |
| MF5 | 0.873 | PS3 | 0.829 |
| MF6 | 0.814 | PS4 | 0.810 |

#### 4.1.1.2. Internal Consistency and Convergent Validity

Internal consistency and convergent validity of the reflective first-order constructs were assessed using CR and AVE. All multi-item reflective constructs met the recommended thresholds outlined in the methodology, with CR values exceeding 0.70 and AVE values above 0.50, indicating acceptable levels of reliability and convergent validity. For the single-item constructs—autonomy and feedback—internal consistency measures are not applicable; instead, their adequacy was evaluated based on their standardized loadings and content validity (Table 4).

*Table 4. Construct reliability and convergent validity (reflective first-order constructs).*

| Construct | CR | AVE |
|---|---|---|
| Meaningfulness | 0.925 | 0.674 |
| Collaboration | 0.758 | 0.512 |
| Pain Points | 0.871 | 0.629 |
| Overall-JS | 0.891 | 0.733 |
| Autonomy (1 item) | n/a | n/a |
| Feedback (1 item) | n/a | n/a |

### 4.1.1.3. Discriminant Validity

Discriminant validity was assessed using both the Fornell–Larcker criterion and HTMT ratio. According to the Fornell–Larcker criterion, the square root of the AVE for each construct was greater than its highest correlation with any other construct, indicating adequate discriminant validity. Specifically, the square root of AVE for BIM-JS (0.800) exceeded its correlation with Overall-JS (0.394), and the square root of AVE for Overall-JS (0.861) similarly exceeded its correlation with BIM-JS. Additionally, HTMT values were all well below the conservative threshold of 0.85, with the highest observed HTMT being 0.395. These results confirm that the latent constructs in the model are empirically distinct from one another.

### 4.1.2. Stage 2: Formative Second-Order Construct (BIM-JS)

### 4.1.2.1. Multicollinearity

VIF was examined to assess collinearity among the formative first-order constructs forming the second-order BIM-JS construct. All VIF values were below the recommended threshold of 3.3, indicating no critical collinearity issues (meaningfulness = 2.12; collaboration = 2.58; autonomy = 1.46; feedback = 1.66; pain points = 2.36).

### 4.1.2.2. Formative Weights Significance (Bootstrapping)

Outer weights of the formative indicators were assessed using bootstrapping with 5,000 resamples. Most formative dimensions (Meaningfulness, Collaboration, Feedback, and Pain Points) showed significant outer weights at $p < .05$, confirming their contribution to the second-order BIM-JS construct. Autonomy did not exhibit a significant outer weight ($p = .274$), suggesting a weaker contribution. In line with recommended practice, outer loadings were also considered when interpreting the substantive relevance of this dimension (Table 5).

*Table 5. Formative Weights and Significance for the BIM–JS Higher-Order Construct*

| First-Order Dimension | Outer Weight | t-value | p-value | Significance |
|---|---|---|---|---|
| Meaningfulness | 0.312 | 6.148 | < .001 | Significant |
| Feedback | 0.378 | 2.418 | .016 | Significant |
| Autonomy | 0.089 | 1.030 | .274 | Not significant |
| Collaboration | 0.385 | 4.716 | < .001 | Significant |
| Pain Points | 0.232 | 4.778 | < .001 | Significant |

## 4.2. Structural Model Evaluation

### 4.2.1. Model Fit, R², f², and Q²

Global model fit was acceptable, with an SRMR of 0.089. The model explained a modest yet meaningful share of variance in Overall Job Satisfaction ($R^2 = 0.163$; $f^2 = 0.195$, medium) and exhibited medium predictive relevance with $Q^2 = 0.175$. Collectively, these findings indicate credible explanatory power and out-of-sample performance.

Overall, the measurement model is statistically sound and suitable for further analysis. Reflective items showed strong reliability, and the constructs demonstrated adequate internal consistency, convergent validity, and discriminant validity. The formative assessment also indicated that each first-order dimension contributed unique information to the second-order BIM-JS construct without problematic collinearity. Together, these results confirm that the model meets key quality criteria and is appropriate for subsequent hypothesis testing.

## 4.3. Hypothesis Testing

In the analysis for RQ1/H1, we examined which work factors contribute most to BIM-JS. Four factors—Collaboration, Meaningfulness, Pain Points, and Feedback—had significant effects, indicating that each plays an important role. Collaboration had the strongest effect (weight = 0.385), followed by Feedback (0.378), Meaningfulness (0.312), and Pain Points (0.232). Autonomy had the lowest weight (0.089) and was not statistically significant at the $\alpha = .05$ level. These results support H1a, H1b, H1d, and H1e, but not H1c (Table 6).

*Table 6. Hypothesis testing (H1)*

| Hypothesis | Dimension | Outer weight | t-value | p-value | Decision |
|---|---|---|---|---|---|
| H1a | Meaningfulness | 0.312 | 6.148 | < .001 | Supported |
| H1b | Feedback | 0.378 | 2.418 | .016 | Supported |
| H1c | Autonomy | 0.089 | 1.030 | .274 | Not supported |
| H1d | Collaboration | 0.385 | 4.716 | < .001 | Supported |
| H1e | Pain Points | 0.232 | 4.778 | < .001 | Supported |

For RQ2/H2, we tested whether BIM-JS significantly predicts Overall Job Satisfaction. The structural path was positive and statistically significant ($\beta = 0.394$, t = 5.159, p < .001), and the bias-corrected 95% confidence interval [0.296, 0.583] did not cross zero, confirming the robustness of this effect. BIM-JS explained 16.3% of the variance in Overall-JS ($R^2 = 0.163$), with a medium effect size ($f^2 = 0.195$). Predictive relevance was also supported ($Q^2 = 0.175$), indicating that the model performs well in terms of out-of-sample prediction. These findings support H2 (Table 7).

*Table 7. Hypothesis Testing (H2, H3)*

| Hypothesis | Path | β | t-value | p-value | Result |
|---|---|---|---|---|---|
| H2 | BIM-JS → Overall-JS | 0.394 | 5.159 | < 0.001 | Supported |
| H3a | Frequency → BIM-JS | 0.336 | 2.53 | <0.05 | Supported |
| H3b | Frequency → Overall-JS | 0.064 | 0.53 | >0.05 | Not Supported |
| H3c | Frequency → BIM-JS → Overall-JS | 0.126 | 2.09 | <0.05 | Supported |

It was also tested whether the frequency of BIM use affects job satisfaction. BIM use frequency significantly predicted BIM-related job satisfaction (BIM-JS), and BIM-JS, in turn, significantly predicted overall job satisfaction (Overall-JS) (Fig. 3). However, BIM use frequency did not have a significant direct effect on Overall-JS. Bootstrapping of the indirect path (Frequency → BIM-JS → Overall-JS) indicated a statistically significant mediation effect ($\beta = 0.126$, 95% bootstrap CI did not include zero), supporting the hypothesized indirect pathway through BIM-JS (Table 7). The model explained 11.3% of the variance in BIM-JS and 16.3% of the variance in Overall-JS. Overall, these findings suggest that simply using BIM more frequently does not, by itself, improve overall job satisfaction; rather, frequency of use enhances job satisfaction primarily when it is associated with more positive BIM-related work experiences (Table 7).

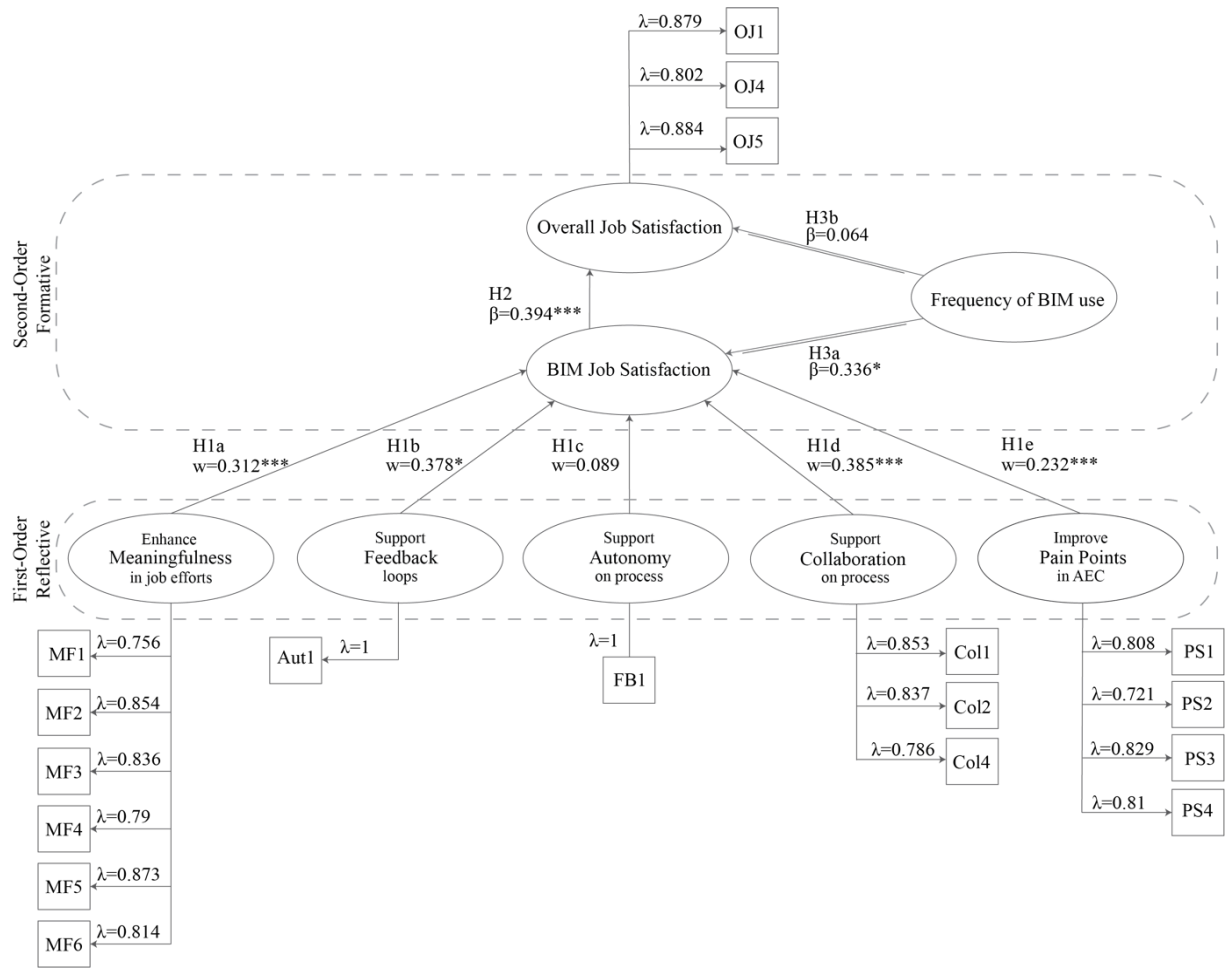

Fig. 3. SEM results. Significance levels: * p < .05, ** p < .01, *** p < .001

## 5. DISCUSSION

This study investigated how BIM-enabled work practices shape job satisfaction in the AEC sector. The results provide strong evidence that BIM influences employee satisfaction primarily through improvements in collaboration, meaningfulness, feedback, and the reduction of pain points in daily tasks. Also, the findings suggest that the impact of BIM depends less on frequency of use and more on how it transforms day-to-day work. The following discussion explores these outcomes in detail and considers their implications for both practice and future research.

### 5.1. RQ1: BIM Factors' Influence on BIM-JS

BIM's role in enhancing meaningfulness was a key contributor to BIM-JS. Professionals reported greater satisfaction when BIM enabled them to contribute visibly to team and client outcomes, expand their skills, and reinforce a sense of purpose. This alignment between personal development and collective impact appears to deepen satisfaction. Organizations seeking to support this factor should position BIM not merely as a technical tool but as a purpose-driven system that enables professionals to acquire new knowledge, apply their expertise, and ultimately observe the tangible outcomes of their efforts. This aligns with previous studies suggesting that when work technologies are designed to enhance job meaning and support intrinsic work values, they can increase engagement and satisfaction (Laschke et al., 2020).

Feedback showed a strong and positive effect on BIM-JS. This suggests that BIM is more satisfying when it helps professionals see the consequences of their decisions and receive timely responses from collaborators or downstream users. This aligns with the idea that digital models are not just design tools but communication and learning tools—when model updates, clash detection, and visualizations provide clear feedback loops, users feel more in control and more aware of their performance. The magnitude of the weight suggests that feedback is nearly as influential as collaboration and meaningfulness, highlighting the importance of using BIM processes and platforms to make outcomes and errors visible in a constructive way.

Autonomy emerged as the only non-significant predictor of BIM–JS. Although BIM technologies offer extensive data and decision-support capabilities, the results indicate that autonomy does not have a substantial impact on job satisfaction in this context. One possible explanation is that shared decision-making is essential for maintaining project coordination in BIM-enabled workflows, which may reduce the perceived importance of individual autonomy. Notably, this perceived limitation did not negatively affect overall job satisfaction. Instead, the findings suggest that in BIM-driven environments, professionals derive greater fulfillment from collaboration than from independent control, possibly highlighting the inherently social nature of BIM-enabled work practices.

Among the factors of BIM-JS, collaboration emerged as the most influential driver. BIM's inherent collaborative features can foster information sharing, joint problem-solving, and coordinated workflows; even modest improvements in perceived collaboration were linked to meaningfully higher job satisfaction. This underscores that to enhance job satisfaction, it is important to cultivate internal and external collaboration. This aligns with de Souza et al (2023), who emphasized that BIM workflows inherently depend on collaboration, making it essential to foster a cooperative environment consistent with AEC industry work patterns. Similarly, another study highlights that insufficient support for collaborative teamwork can hinder effective BIM adoption (Kapogiannis & Sherratt, 2018). The findings of this research reaffirm that collaboration is not merely a byproduct of BIM, but a central mechanism through which BIM enhances work experience.

Pain point reduction, while a more moderate contributor, was still a significant predictor of satisfaction. The responses to survey questions specific to pain points indicated BIM's ability to minimize rework, reduce manual tasks, support smoother workflows, and lower stress from tight deadlines. These findings highlight the importance of intuitive, user-centered design in BIM tools for reducing pain points. This suggests that organizations seeking to enhance job satisfaction should not only promote advanced BIM features but also utilize BIM to reduce reductions in these day-to-day pain points, reinforcing the perception that BIM is solving real problems rather than adding extra workload.

### 5.2. RQ2: BIM-JS Influence on Overall-JS

In response to the second research question, this study examined how BIM-JS influences Overall-JS in the AEC industry. The findings indicate a moderate, positive relationship between BIM-JS and Overall-JS. In this study, higher scores on BIM-related meaningfulness, feedback, collaboration, and reduced pain points were associated with higher BIM-JS, which in turn significantly predicted Overall Job Satisfaction (Tables 5–7; Fig. 3). This suggests that when BIM enhances core aspects of work, such as task clarity, feedback, collaboration, and smoother workflows, professionals report greater satisfaction.

### 5.3. RQ3: Frequency of BIM Use and Overall-JS

In addition to these factor-level insights, the mediation analysis further clarifies how BIM use translates into satisfaction. The effect of BIM use on Overall-JS operated primarily through BIM-JS: more frequent BIM use significantly increased BIM-related job satisfaction, and BIM-JS in turn significantly predicted Overall-JS, while the direct effect of BIM use frequency on Overall-JS was not significant. This pattern indicates that using BIM more often does not, by itself, improve overall job satisfaction unless it is accompanied by more positive BIM-related work experiences. Organizations should therefore prioritize not merely expanding BIM usage, but ensuring that BIM is implemented in ways that enhance day-to-day work quality and satisfaction. Doing so can promote sustained engagement and job fulfillment, reinforcing BIM's value as a human-centered tool for organizational growth.

Taken together, these results indicate that the value of BIM extends beyond its technical functions to the quality of the work experience it enables. Our BIM-JS Measurement Model demonstrates moderate predictive power and theoretical coherence, suggesting that BIM shapes not only how professionals complete their tasks but also how they feel about their work. Positive BIM experiences, especially around collaboration and meaningfulness, appear to elevate broader attitudes and strengthen work enjoyment.

### 5.4. Implications

These findings extend the JCM and related job-satisfaction frameworks into the BIM context, indicating that digital tools are associated with more positive work experiences. Higher perceived meaningfulness and collaboration are key factors of BIM-mediated job satisfaction in contemporary AEC environments. From a managerial perspective, successful BIM adoption depends not only on technical deployment but also on cultivating a collaborative work culture. Our results indicate that satisfaction in BIM-enabled environments is driven more by shared engagement and reduced workflow stress than by greater individual autonomy. Accordingly, leadership should prioritize and create opportunities for enhancing collaboration through fostering teamwork, strengthening both internal and external communication, and facilitating smoother collaborative workflows to enhance overall employee satisfaction. This can be supported through structured mechanisms for cross-disciplinary coordination. Fostering meaningfulness involves linking day-to-day BIM tasks to broader project outcomes and client objectives, helping users see the significance of their work. BIM can also serve as a platform that clarifies the task purpose while enabling knowledge exchange and mutual support among users.

Workflow pain points represent another critical area requiring attention. BIM is designed to reduce friction in project delivery, but its potential is only realized when users are equipped to apply it effectively. Managers should focus on resolving pain points—such as tight deadlines, manual drafting, and design clashes, through the effective use of BIM tools. In parallel, structured feedback systems—such as real-time design review tools or automated progress tracking can enhance learning, recognition, and long-term job enjoyment.

These insights suggest that improving the quality of BIM engagement rather than simply increasing its frequency is key to improving employee satisfaction. Training and implementation strategies should emphasize how BIM enables collaboration, supports feedback, and clarifies task ownership. Encouraging professionals to build a more meaningful relationship with technology can deepen their sense of purpose and motivation. From a theoretical standpoint, these findings reinforce the importance of considering human-centered factors, not just adoption rates, when evaluating digital transformation outcomes. In the context of the AEC industry, BIM should be understood not only as a project management tool but also as a platform for shaping professional experiences, satisfaction, and long-term workforce engagement.

### 5.5. Limitation and Future Direction

While this study offers meaningful insights into how BIM-related work experiences influence job satisfaction, several limitations should be considered when interpreting the results. The sample size (n = 86), though adequate for PLS-SEM, was relatively small and included both public and private sector professionals. A larger sample would strengthen the generalization of the findings and allow for results to be disaggregated by the work sector. Future research should consider analyzing responses from public and private sectors separately to explore how organizational context may influence the relationship between BIM and job satisfaction.

Because the study relied on cross-sectional, self-reported data, it cannot establish causal relationships with certainty. Although the BIM-JS Measurement Model demonstrated good predictive validity, the directionality of the associations was guided by theory rather than confirmed through empirical testing. Longitudinal or quasi-experimental designs could provide a more robust understanding of how BIM-JS develops over time. Qualitative approaches, such as interviews or case studies, would also help contextualize these dynamics and uncover nuanced user experiences. Some factors, most notably autonomy and feedback, were measured using single-item indicators. Although they were theoretically grounded, these measures may have limited the depth and reliability of the analysis. Future research should refine and expand these factors using validated multi-item scales to better capture their complexity and impact.

Finally, the study focused on professionals working in the United States. BIM implementation and its impact on job satisfaction may vary across cultural, regulatory, and organizational contexts. Comparative studies across different countries or regions would provide valuable insights into how local factors shape digital adoption and its human-centered outcomes in the AEC sector. Taken together, these considerations highlight the need for broader, context-sensitive, and methodologically diverse research to deepen and extend the current findings.

### 6. CONCLUSION

This study underscores the significant role of BIM not only as a technical innovation but as a factor in shaping positive work experiences in the AEC industry. By integrating Hackman and Oldham's JCM with AEC-specific challenges, the proposed BIM-JS Measurement Model reveals that collaboration, meaningfulness, feedback about job performance, and the alleviation of workplace pain points are critical drivers of job satisfaction in BIM-enabled environments. However, autonomy, while traditionally emphasized in job satisfaction models, did not show a significant effect, suggesting that BIM's collaborative allowances may reframe the value of independent control in favor of shared coordination. The findings affirm that job satisfaction in digital workspaces hinges less on the frequency of technology use and more on the quality of engagement it fosters, particularly in terms of team collaboration and reduced workflow friction.

More broadly, the results support a human-centric approach to digital transformation in construction, where technological tools are evaluated not solely for their productivity gains but for their impact on well-being and professional fulfillment. For industry leaders and practitioners, this highlights the importance of implementing BIM with a focus on training, team communication, and purposeful design rather than relying on adoption alone. Future research should build upon these insights by incorporating larger, more diverse samples and longitudinal data to better capture the evolving relationship between digital tools and human experience in construction. Ultimately, placing people at the center of digital innovation remains essential for creating sustainable, satisfying, and resilient workplaces in the AEC sector.